\documentclass[reprint,aps,prl,superscriptaddress]{revtex4-2}
\usepackage{amsmath,amssymb,amsfonts}
\usepackage{mathtools}
\usepackage{physics}
\usepackage{graphicx}
\usepackage{tikz}
\usepackage{tikz}
\usetikzlibrary{shapes.geometric, arrows, positioning, fit, backgrounds}
\usepackage{hyperref}

\newcommand{\bfr}{\mathbf{r}}
\newcommand{\bfrp}{\mathbf{r}'}
\newcommand{\rhoh}{\hat{\varrho}}
\newcommand{\HKS}{\hat{H}_{\mathrm{KS}}}
\newcommand{\veff}{v_{\mathrm{eff}}}

\newcommand{\calL}{\mathcal{L}}
\newcommand{\calN}{\mathcal{N}}
\newcommand{\calZ}{\mathcal{Z}}
\newcommand{\kB}{k_B}

\tikzstyle{process} = [rectangle, minimum width=3.5cm, minimum height=1cm, text centered, draw=black, fill=blue!5]
\tikzstyle{decision} = [diamond, minimum width=3cm, minimum height=1cm, text centered, draw=black, fill=green!5]
\tikzstyle{io} = [trapezium, trapezium left angle=70, trapezium right angle=110, minimum width=3cm, minimum height=1cm, text centered, draw=black, fill=orange!5]
\tikzstyle{arrow} = [thick,->,>=stealth]

\begin{document}

\title{Fundamental Relation between Conductance of Biomolecules and the Fukui Function}

\author{G\'abor Vattay}
\email{gabor.vattay@ttk.elte.hu}
\affiliation{Institute for Physics and Astronomy, E\"otv\"os University, H-1053 Budapest, Egyetem t\'er 1-3., Hungary}

\begin{abstract}
We derive the finite-temperature conductance of a molecule coupled to metallic leads using open-system time-dependent density functional theory (TDDFT). Tracing out the phonon bath via Wick's theorem yields an exchange-symmetric collision integral that preserves Pauli exclusion. When internal thermalization is faster than lead escape---as holds in proteins with multifractal orbitals at the Anderson transition---the exchange-correlation kernel projects out of the slow Liouvillian mode, making conductance directly proportional to the finite-temperature Fukui function of conceptual DFT.
\end{abstract}

\maketitle

Experiments in protein bioelectronics have shown that single proteins and peptide monolayers conduct electrons over nanometer distances far more efficiently than expected for conventional organic insulators~\cite{Amdursky2014, SepunaruKronik2015, GuoKronik2016, Bostick2018, Fereiro2018, Zhang2019, Zhang2020JACS, Papp2023}. Scanning tunneling microscope break-junction measurements reveal nanosiemens-scale conductance through single folded proteins in aqueous solution, governed primarily by the electrode-protein contact chemistry rather than the molecular length~\cite{Zhang2019, Zhang2020JACS}. Solid-state junctions display thermally activated transport above roughly $150$~K and cross over to temperature-independent transport at lower temperatures~\cite{Amdursky2014, Fereiro2018, Papp2023}. Because long-range conduction occurs across both redox and non-redox proteins and homopeptides~\cite{SepunaruKronik2015, GuoKronik2016, Bostick2018, Zhang2020JACS}, its physical origin must lie in the electronic structure of the folded polypeptide scaffold itself.

A physical mechanism for this behavior comes from quantum criticality at the Anderson metal-insulator transition~\cite{EversMirlin2008, VattayKauffman2014, PappVattay2025}. The electronic Hamiltonians of naturally occurring proteins sit near the critical point of the Anderson transition and exhibit critical level statistics at the edge of quantum chaos~\cite{VattayKauffman2014}. Their Kohn-Sham (KS) wave functions are multifractal, extending across the molecular volume with self-similar spatial fluctuations~\cite{EversMirlin2008, PappVattay2025}. Near the Fermi level, the highest occupied and lowest unoccupied molecular orbitals (HOMO and LUMO) have a correlation fractal dimension $d_2 \approx 2$, so that orbital probability scales effectively with the protein surface area~\cite{PappVattay2025}. This extended multifractal structure enables rapid anomalous diffusion of injected charges across the protein scaffold, keeping the internal electron-phonon relaxation time $\tau_{\mathrm{rlx}}$ much shorter than the escape time $\tau_{\mathrm{esc}}$ into the metallic electrodes. In this fast-thermalization regime, the finite-temperature linear conductance takes the symmetric form~\cite{PappVattay2024}:
\begin{equation}
\label{eq:GT_intro}
G_T = \frac{2e^2}{\hbar}\,\frac{Z_L\,Z_R}{Z_L + Z_R},
\end{equation}
where $Z_L$ and $Z_R$ characterize the local electronic structure at the left and right electrode contacts.

In conceptual density functional theory (CDFT), the Fukui function $f(\bfr) = (\partial \rho(\bfr)/\partial N)_{v}$ is the primary local reactivity descriptor~\cite{ParrYang1984, YangParr1985, ParrYangBook1989, Geerlings2003}. It identifies molecular sites susceptible to nucleophilic or electrophilic attack and maps enzymatic active sites and allosteric pockets in biological macromolecules~\cite{Roos2009}. Extended to finite temperatures in the grand canonical ensemble, the Fukui function incorporates thermal Fermi-Dirac broadening across the frontier orbitals~\cite{FrancoPerez2015b}.

In this Letter, we show that for any open molecular junction satisfying the fast-thermalization condition, the spatial transport profile $\calZ(\bfr)$ is proportional to the finite-temperature Fukui function, $\calZ(\bfr) = \calN f_T(\bfr)$. Working within time-dependent DFT for open quantum systems, we prove that the dynamical Hartree and exchange-correlation self-consistent field response is projected out of the steady-state transport mode. Whereas exchange-correlation kernels introduce nontrivial corrections to coherent Landauer-Kohn-Sham conductance~\cite{BurkeCarGebauer2005}, fast internal equilibration decouples them here, uniting molecular transport theory with conceptual DFT.

By the Hohenberg-Kohn theorem~\cite{HohenbergKohn1964}, the ground-state energy of an interacting electron system in an external potential $v(\bfr)$ is a functional of the electron density $\rho(\bfr)$:
\begin{equation}
\label{eq:E_functional}
E[\rho] = T_s[\rho] + \int v(\bfr)\rho(\bfr)d\bfr + E_H[\rho] + E_{xc}[\rho],
\end{equation}
where $T_s[\rho]$ is the non-interacting kinetic energy, $E_H[\rho] = \frac{1}{2}\iint \rho(\bfr)\rho(\bfrp)|\bfr-\bfrp|^{-1}d\bfr d\bfrp$ is the Hartree electrostatic energy, and $E_{xc}[\rho]$ is the exchange-correlation functional. At finite temperature $T$ and chemical potential $\mu$, minimizing the Mermin grand potential functional~\cite{Mermin1965}, $\Omega[\rho; T, \mu] = E[\rho] - T S_s[\rho; T] - \mu N[\rho]$, yields the self-consistent Kohn-Sham eigenvalue equations~\cite{KohnSham1965}:
\begin{equation}
\label{eq:KS_eq}
\left( -\frac{\hbar^2}{2m}\nabla^2 + v_{\mathrm{eff}}[\rho](\bfr) \right) \psi_k(\bfr) = E_k \psi_k(\bfr),
\end{equation}
with the effective potential
\begin{equation}
\label{eq:veff}
v_{\mathrm{eff}}[\rho](\bfr) = v(\bfr) + \int \frac{\rho(\bfrp)}{|\bfr - \bfrp|}d\bfrp + \frac{\delta E_{xc}}{\delta \rho(\bfr)}.
\end{equation}
The equilibrium electron density is
\begin{equation}
\label{eq:rho_eq}
\rho(\bfr) = \sum_k F(E_k, \mu) |\psi_k(\bfr)|^2,
\end{equation}
where $F_k \equiv F(E_k, \mu) = [1 + \exp((E_k - \mu)/\kB T)]^{-1}$ is the Fermi-Dirac occupancy and $\rhoh_{\mathrm{eq}} = \sum_k F_k \ket{\psi_k}\bra{\psi_k}$ is the equilibrium single-particle KS density matrix.

The chemical potential $\mu = (\partial E/\partial N)_v$ and the density $\rho(\bfr)$ obey the Maxwell relation $(\partial \rho(\bfr)/\partial N)_{v} = (\delta \mu/\delta v(\bfr))_{N}$~\cite{ParrYangBook1989}, which defines the zero-temperature Fukui function from the energy functional~\cite{ParrYang1984, YangParr1985}. In the finite-temperature grand canonical ensemble, the Fukui function is given by $f_T(\bfr) = (\partial\rho(\bfr)/\partial\mu)_{v,T} / (\partial N/\partial\mu)_{v,T}$~\cite{FrancoPerez2015b}. Within the frozen-orbital approximation of CDFT, where the KS orbitals $\psi_k(\bfr)$ are held fixed under infinitesimal shifts of $\mu$ while their thermal occupations vary, the density derivative is
\begin{equation}
\label{eq:drho_dmu}
\frac{\partial \rho(\bfr)}{\partial \mu} = \sum_k \frac{|\psi_k(\bfr)|^2}{4\kB T \cosh^2\!\left(\frac{E_k - \mu}{2\kB T}\right)}.
\end{equation}
Defining the dimensionless normalization factor $\calN$ related to the global softness $S = (\partial N/\partial\mu)_{v,T} = \calN/(4\kB T)$ as
\begin{equation}
\label{eq:calN_def}
\calN = \sum_k \frac{1}{\cosh^2\!\left(\frac{E_k - \mu}{2\kB T}\right)},
\end{equation}
we obtain the finite-temperature Fukui function:
\begin{equation}
\label{eq:fukui_T}
f_T(\bfr) = \frac{1}{\calN} \sum_k \frac{|\psi_k(\bfr)|^2}{\cosh^2\!\left(\frac{E_k - \mu}{2\kB T}\right)}.
\end{equation}

To describe non-equilibrium transport, we use open-system time-dependent DFT (OQS-TDDFT)~\cite{RungeGross1984, BurkeCarGebauer2005, YuenZhou2010}. The open-system KS single-particle density matrix $\rhoh$ obeys the master equation
\begin{equation}
\label{eq:OQS_KS}
\partial_t\,\rhoh = \frac{1}{i\hbar}\bigl[\HKS[\rho],\,\rhoh\bigr] + R(\rhoh) + \mathcal{D}_{\mathrm{leads}}(\rhoh),
\end{equation}
where $R(\rhoh)$ describes relaxation due to the vibrational phonon bath and $\mathcal{D}_{\mathrm{leads}}(\rhoh)$ accounts for electron exchange with the metallic leads. To obtain $R(\rhoh)$ within the KS reference system, we treat the molecular electrons and vibrational modes as a composite system before tracing out the bath:
\begin{equation}
\label{eq:Htot}
\hat{H}_{\mathrm{tot}} = \HKS[\rho] + \sum_\alpha \hbar\omega_\alpha \hat{b}_\alpha^\dagger\hat{b}_\alpha + \sum_\alpha g_\alpha \hat{\rho}(\bfr_\alpha) (\hat{b}_\alpha + \hat{b}_\alpha^\dagger).
\end{equation}
Here $\hat{H}_B = \sum_\alpha \hbar\omega_\alpha \hat{b}_\alpha^\dagger \hat{b}_\alpha$ is the free phonon Hamiltonian and $\hat{H}_{eB} = \sum_\alpha g_\alpha \hat{\rho}(\bfr_\alpha)(\hat{b}_\alpha + \hat{b}_\alpha^\dagger)$ couples the KS density operator $\hat{\rho}(\bfr) = \hat{\Psi}^\dagger(\bfr)\hat{\Psi}(\bfr)$ linearly to the local phonon displacement with strength $g_\alpha$. Because electronic exchange and correlation are already represented by $\veff[\rho]$, the electron-phonon coupling acts directly on the non-interacting KS fermions.

Under the Born-Markov approximation ($\rhoh_{\mathrm{tot}} \approx \rhoh \otimes \rhoh_B$ with rapid bath memory decay)~\cite{BreuerPetruccione2002}, tracing out the phonon modes yields the Redfield equation containing two-time four-fermion correlation functions of the KS creation ($\hat{c}_i^\dagger$) and annihilation ($\hat{c}_j$) operators. Because the KS reference system is non-interacting, Wick's theorem factorizes these correlations into single-particle density matrix elements $\varrho_{ij} = \langle \hat{c}_j^\dagger \hat{c}_i \rangle$:
\begin{equation}
\label{eq:Wick_main}
\langle \hat{c}_i^\dagger \hat{c}_j \hat{c}_p^\dagger \hat{c}_q \rangle = \varrho_{ji}\,\varrho_{qp} + \varrho_{qi}\,(\delta_{jp} - \varrho_{jp}).
\end{equation}
The second term is the exchange contraction arising from the canonical anticommutation relations $\{\hat{c}_i, \hat{c}_j^\dagger\} = \delta_{ij}$. Retaining this contraction enforces Pauli blocking at the full operator matrix level via $(\hat{I} - \rhoh)$ rather than merely through scalar diagonal occupations $\varrho_{pp}(1 - \varrho_{pp})$, yielding the exchange-symmetric collision integral
\begin{equation}
\label{eq:R_anticomm}
R(\rhoh) = \tfrac{1}{2}\bigl\{\hat{\Sigma}^<(\rhoh),\,\hat{I} - \rhoh\bigr\} - \tfrac{1}{2}\bigl\{\hat{\Sigma}^>(\rhoh),\,\rhoh\bigr\},
\end{equation}
where $\{\cdot,\cdot\}$ is the anticommutator, and the lesser and greater electron-phonon self-energies are $\Sigma^<_{ij}(\rhoh) = \sum_{pq} W^<_{iq,pj}\varrho_{qp}$ and $\Sigma^>_{ij}(\rhoh) = \sum_{pq} W^>_{iq,pj}(\delta_{qp} - \varrho_{qp})$. In terms of the KS electron-phonon matrix elements $\rho_\alpha^{ip} \equiv \langle\psi_i|\hat{\rho}_\alpha|\psi_p\rangle = \int \psi_i^*(\bfr)\rho_\alpha(\bfr)\psi_p(\bfr)d\bfr$ and the Bose-Einstein distribution $n_B(\omega) = (\exp(\hbar\omega/\kB T) - 1)^{-1}$ at transition frequency $\omega_{pq} = (E_p - E_q)/\hbar$, the transition rate tensors take the form
\begin{align}
\label{eq:W_main}
W^<_{iq,pj} &= \frac{2\pi}{\hbar}\sum_\alpha |g_\alpha|^2\,\rho_\alpha^{ip}\rho_\alpha^{qj}\,n_B(\omega_{pq})\,\delta(\omega_{pq} - \omega_\alpha),\nonumber\\
W^>_{iq,pj} &= \frac{2\pi}{\hbar}\sum_\alpha |g_\alpha|^2\,\rho_\alpha^{ip}\rho_\alpha^{qj}\,[1 + n_B(\omega_{pq})]\,\delta(\omega_{pq} - \omega_\alpha).
\end{align}
These tensors obey detailed balance, $W^<_{iq,pj}/W^>_{pj,iq} = \exp(-\hbar\omega_{pq}/\kB T)$. In the KS eigenbasis, the matrix elements of $R(\rhoh)$ read
\begin{align}
\label{eq:Rij_explicit}
R^{ij}(\rhoh) = \frac{1}{2}\sum_k&\left[\Sigma^<_{ik}(\delta_{kj} - \varrho_{kj}) + (\delta_{ik} - \varrho_{ik})\Sigma^<_{kj}\right.\\ \nonumber
&- \left.\Sigma^>_{ik}\varrho_{kj} - \varrho_{ik}\Sigma^>_{kj}\right],
\end{align}
which vanishes identically at the Mermin equilibrium state, $R(\rhoh_{\mathrm{eq}}) = 0$ (see Sec.~S1 of the Supplemental Material~\cite{SM}).

We next couple the molecular contact sites $n \in \{L, R\}$ to metallic electrodes in the wide-band limit~\cite{MeirWingreen1992}. The coupling introduces escape rate matrices $\Gamma^{ij} = \sum_n \psi_n^{i*}\Gamma_n \psi_n^j$ and modifies the master equation via $\mathcal{D}_{\mathrm{leads}}(\rhoh) = -\frac{1}{2\hbar}\{\hat{\Gamma},\,\rhoh\} + \hat{J}$. Integrating the TDDFT continuity equation $\partial_t\rho(\bfr, t) + \nabla\cdot\mathbf{j}_{\mathrm{KS}}(\bfr, t) = \text{source/sink}$ over the molecular volume gives the rate of change of the total electron number $N = \mathrm{Tr}(\rhoh)$:
\begin{equation}
\label{eq:dN_dt_main}
\frac{dN}{dt} = \sum_{n=L,R}\frac{1}{\hbar}\mathrm{Tr}\!\bigl(\hat{\Gamma}_n[\rhoh_{\mathrm{eq}}(\mu_n) - \rhoh]\bigr) + \mathrm{Tr}\bigl(R(\rhoh)\bigr).
\end{equation}
Because electron-phonon scattering conserves particle number on the molecule, $\mathrm{Tr}(R(\rhoh)) = 0$. In the steady state ($dN/dt = 0$), the current entering from the left lead equals that leaving into the right lead, $J_L = -J_R = J$, giving
\begin{equation}
\label{eq:JL}
J_L = \frac{1}{\hbar}\mathrm{Tr}\!\bigl(\hat{\Gamma}_L\,[\rhoh_{\mathrm{eq}}(\mu_L) - \rhoh]\bigr).
\end{equation}

Under a small symmetric bias voltage $U$ ($\mu_{L/R} = \mu \pm eU/2$), the density matrix departs linearly from equilibrium, $\rhoh = \rhoh_{\mathrm{eq}} + \delta\rhoh$. In TDDFT linear response~\cite{RungeGross1984, BurkeCarGebauer2005}, the resulting density shift $\delta\rho(\bfr)$ perturbs the self-consistent KS Hamiltonian:
\begin{equation}
\label{eq:dHKS}
\delta H_{\mathrm{KS}}(\bfr) = \int \left( \frac{1}{|\bfr - \bfrp|} + f_{xc}(\bfr, \bfrp) \right) \delta\rho(\bfrp) d\bfrp,
\end{equation}
where $f_{xc}(\bfr, \bfrp) = \delta^2 E_{xc}/\delta\rho(\bfr)\delta\rho(\bfrp)$ is the exchange-correlation kernel. In the KS eigenbasis, this self-consistent field (SCF) response enters Eq.~\eqref{eq:OQS_KS} through the commutator $[\delta \hat{H}_{\mathrm{KS}}, \rhoh_{\mathrm{eq}}]^{ij} = (F_j - F_i)\sum_{pq}K^{ij,pq}\delta\varrho^{pq}$, where $K^{ij,pq}$ are the two-electron matrix elements of the combined Hartree-XC kernel.

Linearizing Eq.~\eqref{eq:OQS_KS} around $\rhoh_{\mathrm{eq}}$ in the stationary state ($\partial_t\rhoh = 0$) gives
\begin{equation}
\label{eq:linearized_eq}
\sum_{pq}\calL^{ijpq}\,\delta\varrho^{pq} = -\hbar J^{ij},
\end{equation}
with the linearized KS Liouvillian superoperator
\begin{align}
\calL^{ijpq} = &-i(E_i - E_j)\,\delta_{ip}\delta_{jq} - i(F_j - F_i)\,K^{ij,pq}\\ \nonumber
&- \tfrac{1}{2}(\Gamma^{ip}\delta_{jq} + \delta_{ip}\Gamma^{qj}) + \delta R^{ijpq},
\label{eq:calL_def}
\end{align}
and the lead injection driving term $J^{ij} = \frac{eU}{2\hbar}(\Gamma_L^{ij} - \Gamma_R^{ij})[f(E_i, \mu) + f(E_j, \mu)]$, where $f(E, \mu) = \partial_\mu F(E, \mu) = [4\kB T\cosh^2((E - \mu)/2\kB T)]^{-1}$.

For the closed molecule ($\Gamma=0$), particle conservation ensures that the unperturbed superoperator $\calL_0$ has a zero eigenvalue $\lambda_0 = 0$. Its left and right eigenvectors are purely diagonal in the KS energy basis: $U_0^{ij} = \delta_{ij}$ and $V_0^{ij} = \delta_{ij}\calN^{-1}\cosh^{-2}[(E_i - \mu)/2\kB T]$. The spectral gap to the first non-zero eigenvalue $\lambda_1$ sets the internal thermalization time $\tau_{\mathrm{rlx}} = \hbar/|\lambda_1|$.

A simple commutator identity governs the projection onto this slow mode: for any operator $\hat{A}$ and any operator $\hat{B} = \sum_k B_k\ket{k}\bra{k}$ that is diagonal in the KS basis, the diagonal matrix elements of their commutator vanish identically,
\begin{equation}
\label{eq:comm_identity}
[\hat{A}, \hat{B}]^{ii} = \sum_k (A^{ik}B_k\delta_{ki} - B_i\delta_{ik}A^{ki}) = A^{ii}B_i - B_i A^{ii} = 0.
\end{equation}
Setting $\hat{A} = \delta\hat{H}_{\mathrm{KS}}$ and $\hat{B} = \rhoh_{\mathrm{eq}}$ (which is diagonal in the KS eigenbasis) shows immediately that the SCF response term has no diagonal entries:
\begin{equation}
\label{eq:proof_SCF_zero}
L_{\mathrm{SCF}}^{iipq} = -i(F_i - F_i)K^{ii,pq} = 0, \qquad \forall p,q.
\end{equation}
Likewise, the off-diagonal exchange terms in the linearized collision integral $\delta R^{ijpq}$ vanish on the diagonal.

In the fast-thermalization regime ($\tau_{\mathrm{rlx}} \ll \tau_{\mathrm{esc}} \approx \hbar/\Gamma$), off-diagonal coherences $\delta\varrho^{ij}$ ($i \neq j$) decay rapidly before charges escape to the leads. Projecting $\calL$ onto the slow mode via the diagonal eigenvectors $U_0$ and $V_0$ restricts the steady-state transport dynamics to the diagonal population sector, so the Hartree-XC kernel response projects identically to zero:
\begin{equation}
\label{eq:project_zero_main}
\sum_{ij,pq} U_0^{ij}\, L_{\mathrm{SCF}}^{ijpq}\, V_0^{pq} = \sum_{i,pq} L_{\mathrm{SCF}}^{iipq}\, V_0^{pq} = 0.
\end{equation}
This also explains why the frozen-orbital Fukui function emerges naturally: orbital relaxation under a chemical potential shift corresponds to off-diagonal coherent mixing, which belongs to the fast-decaying subspace.

Treating the weak lead coupling $\Gamma$ as a perturbation shifts the zero eigenvalue to first order:
\begin{equation}
\label{eq:lambda0}
\lambda_0 = -\frac{1}{\calN}\sum_k \frac{\Gamma^{kk}}{\cosh^2\!\left(\frac{E_k - \mu}{2\kB T}\right)}.
\end{equation}
Because $|\lambda_0| \ll |\lambda_1|$, the inverse superoperator is dominated by the slow mode, $\calL^{-1} \approx \lambda_0^{-1} V_0 U_0$, giving $\delta\varrho^{pq} \approx -\hbar\lambda_0^{-1}V_0^{pq}\sum_{ij} U_0^{ij} J^{ij}$ (Sec.~S2 of Ref.~\cite{SM}). Substituting this into Eq.~\eqref{eq:JL} yields the conductance formula of Eq.~\eqref{eq:GT_intro}, where the contact parameters are
\begin{equation}
\label{eq:ZLR}
Z_{L/R} = \frac{\Gamma_{L/R}}{4\kB T}\sum_k \frac{|\psi_k(\bfr_{L/R})|^2}{\cosh^2\!\left(\frac{E_k - \mu}{2\kB T}\right)}.
\end{equation}
Defining the spatial conductance profile $\calZ(\bfr) \equiv \sum_k |\psi_k(\bfr)|^2\cosh^{-2}[(E_k - \mu)/2\kB T]$ and comparing it with Eq.~\eqref{eq:fukui_T}, we arrive at our central identity:
\begin{equation}
\label{eq:main_result}
\calZ(\bfr) = \calN\,f_T(\bfr).
\end{equation}
The spatial conductance profile is directly proportional to the finite-temperature Fukui function. Expressing $\calN = 4\kB T\,S$ in terms of the global softness $S$ and the local softness $s_T(\bfr) = S f_T(\bfr)$, the contact parameters take the compact form
\begin{equation}
\label{eq:ZLR_softness}
Z_{L/R} = \Gamma_{L/R}\, S\, f_T(\bfr_{L/R}) = \Gamma_{L/R}\, s_T(\bfr_{L/R}).
\end{equation}
Thus, the finite-temperature conductance of a biomolecule is determined by its local chemical reactivity at the contact sites.

\textit{Timescale separation in quantum critical proteins.}---We can verify quantitatively why folded proteins satisfy the fast-thermalization hierarchy $\tau_{\mathrm{rlx}} \ll \tau_{\mathrm{esc}}$. In the closed protein, the average relaxation gap $|\lambda_1|$ of $\calL_0$ is set by the frequency overlap between the vibrational bath spectral density $\gamma(\omega)$ and the local density-density correlation function $C(\omega)$ of KS orbitals separated by energy $\hbar\omega$:
\begin{equation}
\label{eq:lambda1_main}
\langle |\lambda_1| \rangle = \int_0^\infty \gamma(\omega)\,C(\omega)\,d\omega.
\end{equation}
For an Ohmic protein vibrational bath $\gamma(\omega) = \eta\hbar\omega e^{-\omega/\omega_c}/(e^{\hbar\omega/\kB T} - 1) \approx \eta\kB T e^{-\omega/\omega_c}$ at room temperature ($\kB T \approx 0.025$~eV), typical protein parameters are $\hbar\omega_c \approx 0.0185$~eV and dimensionless coupling $\eta \approx 1.46$~\cite{Mohseni2008}. Because the critical KS orbitals in $d = 3$ dimensions are multifractal with correlation dimension $d_2 \approx 2$, their spatial overlap follows the Chalker scaling $C(\omega) = A\omega^{-(1 - d_2/d)} = A\omega^{-1/3}$ with $A \approx 2.15\text{ eV}^{1/3}$~\cite{Chalker1990, PappVattay2025}. Evaluating Eq.~\eqref{eq:lambda1_main} yields $\langle |\lambda_1| \rangle \approx \eta\kB T A\omega_c^{2/3}\Gamma(2/3) \approx 7.4$~meV, corresponding to a sub-picosecond internal thermalization time $\tau_{\mathrm{rlx}} = \hbar/|\lambda_1| \approx 89$~fs.

By contrast, the escape width $\Gamma_k$ of an extended orbital coupled to a localized electrode contact is diluted by the inverse of the molecular volume: $\Gamma_k = \Gamma_{\mathrm{local}}|\psi_k(\bfr_{\mathrm{contact}})|^2$. Even for a strong chemical bond at the contact ($\Gamma_{\mathrm{local}} \approx 100$~meV), delocalization across a protein of $N \approx 3,000$ atoms reduces the local orbital weight to $|\psi_k(\bfr_{\mathrm{contact}})|^2 \sim 1/N \approx 3\times 10^{-4}$, giving $\Gamma_k \approx 30\ \mu$eV and an escape time $\tau_{\mathrm{esc}} = \hbar/\Gamma_k \approx 22$~ps. The escape rate is thus 250 times smaller than the internal relaxation gap ($\Gamma_k \ll |\lambda_1|$). In an ordinary Anderson insulator with localization decay constant $\beta \approx 0.5\text{ \AA}^{-1}$, orbital overlaps across a $L \approx 30$~\AA~domain would instead be suppressed by $e^{-2\beta L}$, collapsing $|\lambda_1|$ to $\sim 10^{-9}$~eV ($\tau_{\mathrm{rlx}} \sim 1\ \mu\text{s} \gg \tau_{\mathrm{esc}}$) and violating fast thermalization.

\begin{figure}[t]
    \centering
    \includegraphics[width=0.35\columnwidth]{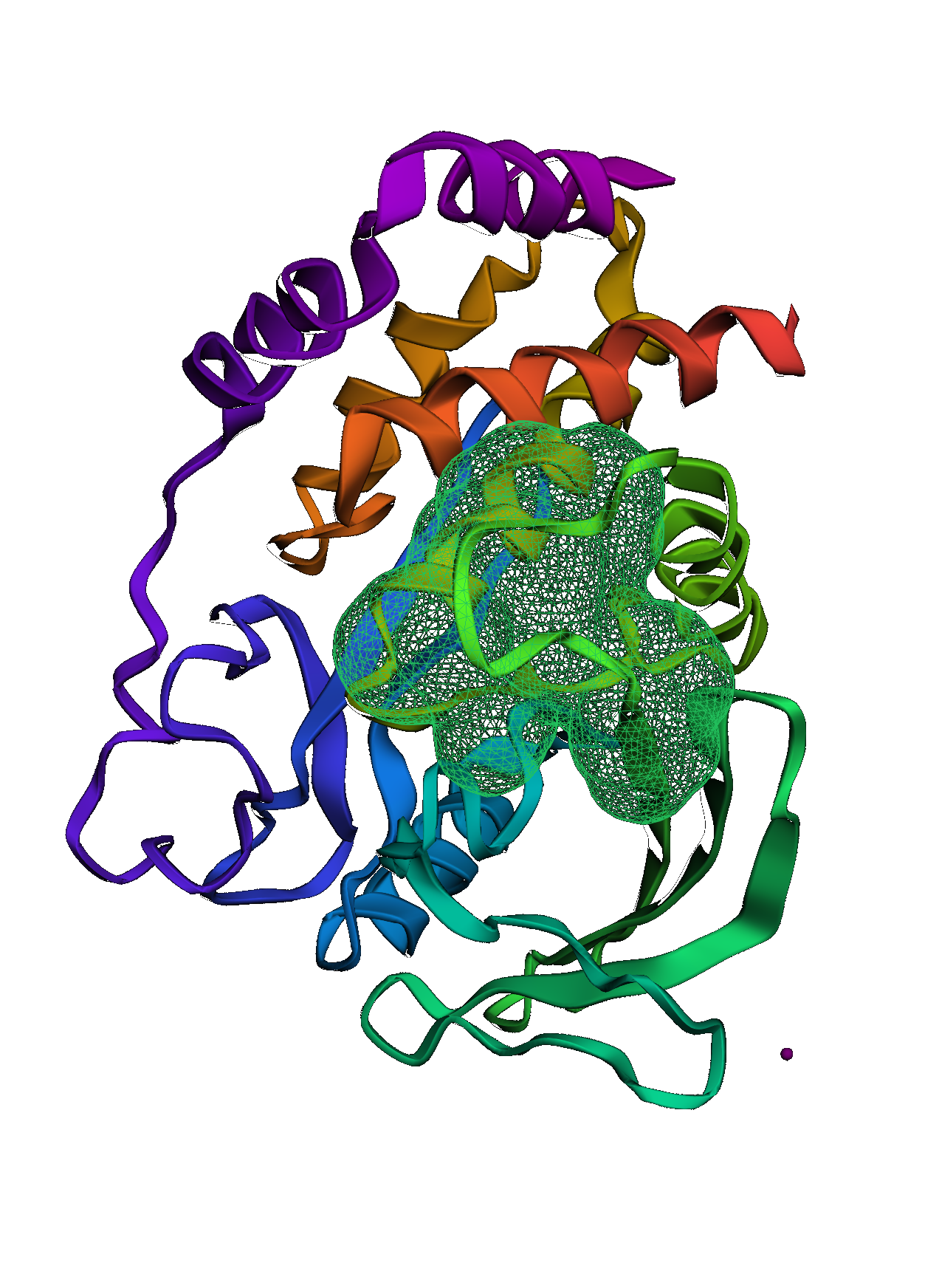} \hfill
    \includegraphics[width=0.59\columnwidth]{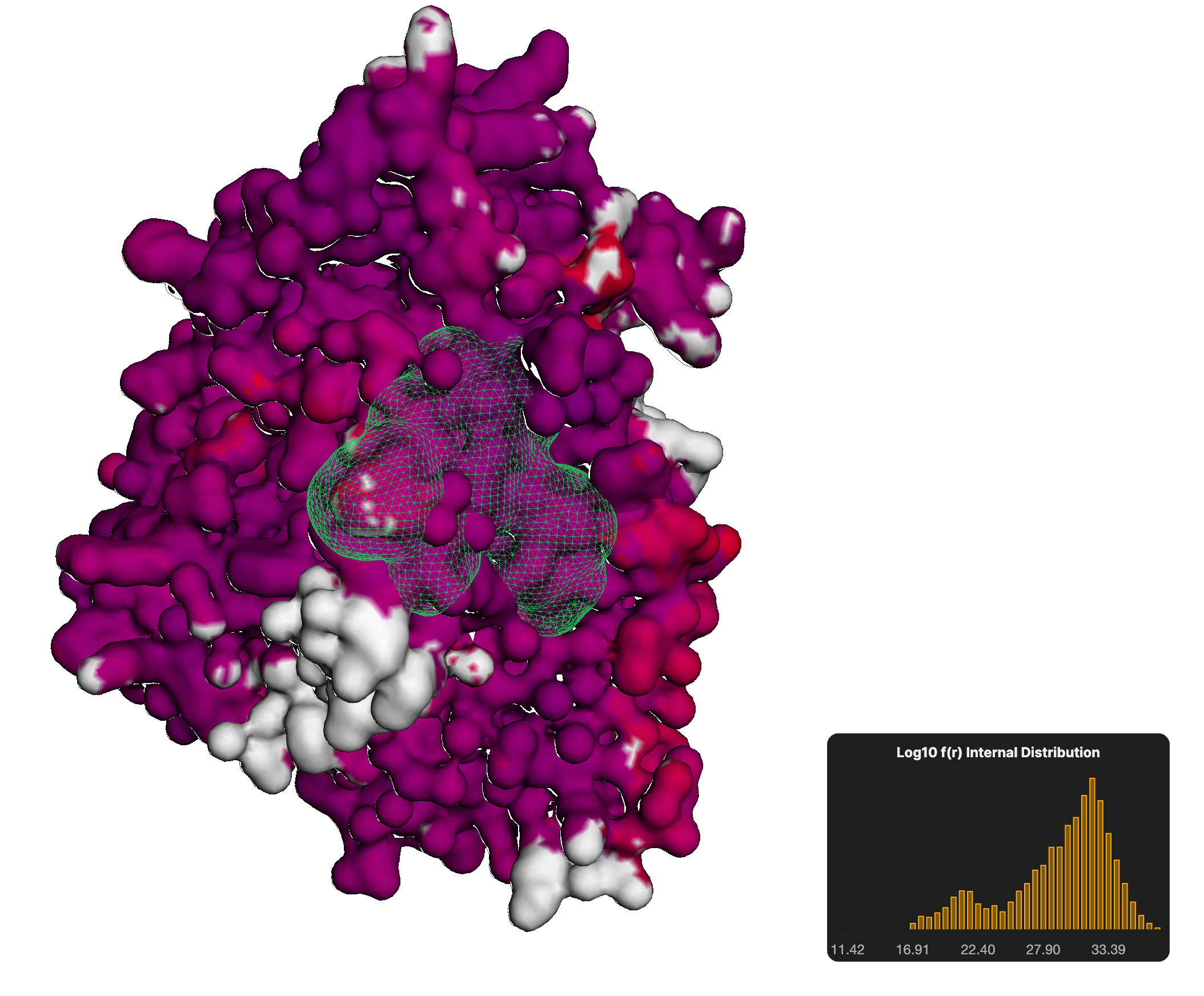} \\
    \vspace{0.1cm}
    \includegraphics[width=0.35\columnwidth]{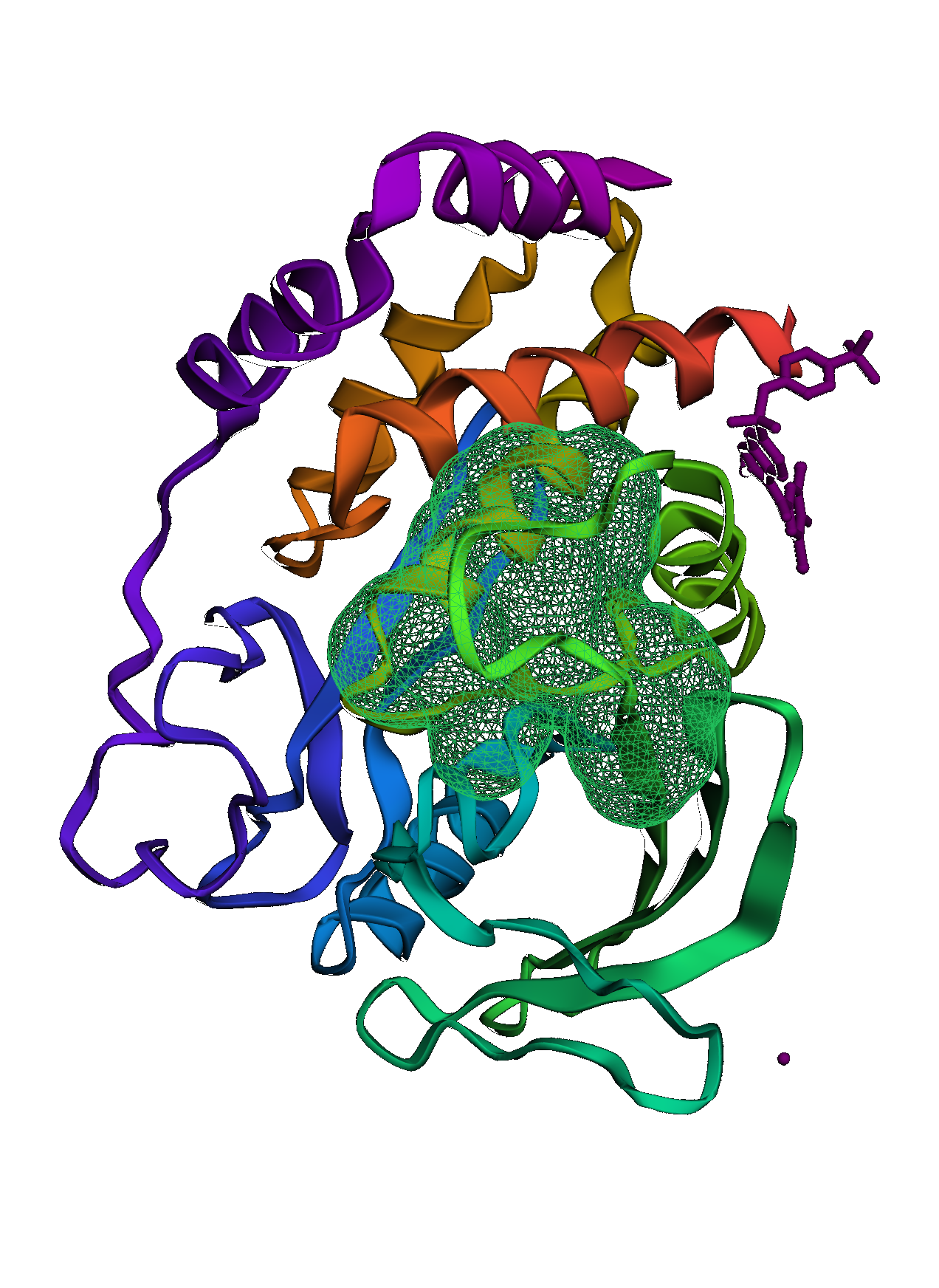} \hfill
    \includegraphics[width=0.59\columnwidth]{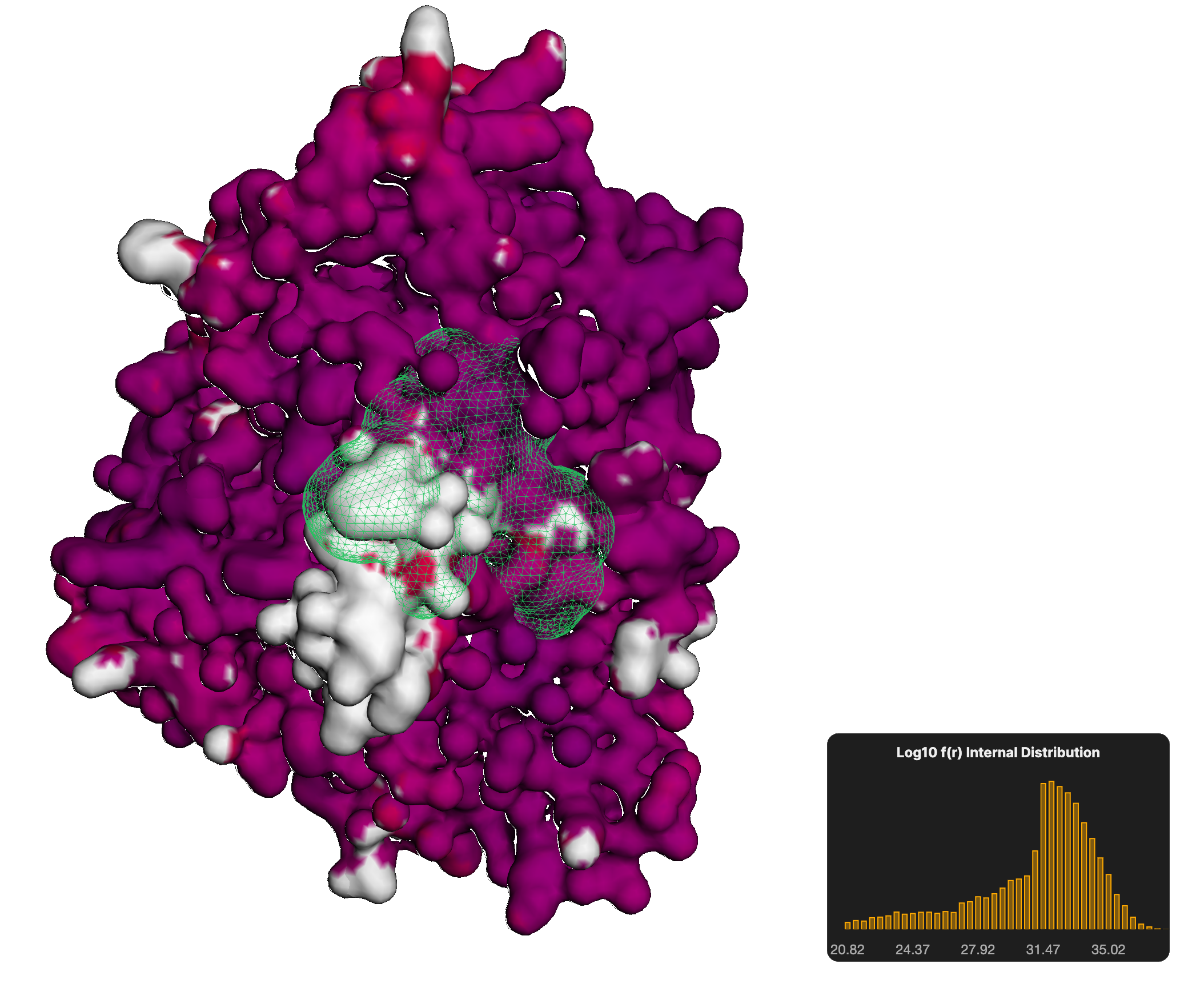}
    \caption{Conductance-reactivity profile in protein tyrosine phosphatase 1B (PTP1B, PDB ID: 1T49). Top row: Apo (ligand-free) structure (left) and the spatial profile of the finite-temperature Fukui function $f_T(\bfr)$ across the catalytic WPD loop (green mesh, right). Bottom row: Holo structure bound to the benzofuran sulfonamide allosteric inhibitor (left) and the resulting Fukui function (right), showing suppression of the conductive reactivity pathway at the WPD loop upon ligand binding. Insets: internal histograms (yellow) of $\ln f_T(\bfr)$ sampled on a uniform grid inside the protein volume (shifted by $+40$ on the horizontal axis), displaying the log-normal distribution characteristic of multifractal wave functions at the Anderson transition.}
    \label{fig:ptp1b_validation}
\end{figure}

\textit{Application to allosteric regulation.}---To illustrate this relation in a biological setting, we examine allosteric inhibition in protein tyrosine phosphatase 1B (PTP1B), a key enzyme in insulin receptor signaling and a therapeutic target for type II diabetes and obesity~\cite{Wiesmann2004}. Catalysis in PTP1B depends on the closure of its conserved Trp-Pro-Asp (WPD) loop over the active site. Binding of a benzofuran sulfonamide inhibitor to an allosteric pocket roughly $20$~\AA~from the catalytic site blocks loop closure and locks the enzyme in its inactive open conformation~\cite{Wiesmann2004}.

According to Eqs.~\eqref{eq:main_result} and~\eqref{eq:ZLR_softness}, allosteric inhibition should manifest as a spatial redistribution of the finite-temperature Fukui function connecting the allosteric pocket to the active site. We computed $f_T(\bfr)$ at $T = 300$~K for PTP1B (PDB ID: 1T49) in both the ligand-free (apo) and inhibitor-bound (holo) structures using self-consistent-charge density-functional tight-binding (SCC-DFTB+)~\cite{Hourahine2020}. After protonating the crystallographic coordinates at physiological pH and relaxing the hydrogen network under heavy-atom harmonic restraints ($k = 10.0\text{ kJ/mol/nm}^2$), we extracted the interior eigenstates around the converged Fermi level ($E_F = -3.4682$~eV) via shift-invert Arnoldi diagonalization and projected the Slater-type orbitals onto a uniform $1.5$~\AA~three-dimensional Cartesian grid (see Sec.~S3 of Ref.~\cite{SM} for the automated structural pipeline).

Figure~\ref{fig:ptp1b_validation} compares the resulting spatial profiles. In the apo enzyme, $f_T(\bfr)$ forms a continuous bridge extending from the allosteric site into the catalytic WPD loop, supporting efficient charge redistribution during loop closure. When the benzofuran sulfonamide ligand binds in the holo complex, $f_T(\bfr)$ is strongly suppressed across the WPD loop and the adjacent allosteric channel. By depleting the local softness along this pathway, the inhibitor disrupts the electronic coupling needed for loop mobility.

\textit{Physical interpretation of conductance.}---Equation~\eqref{eq:GT_intro} corresponds to the series combination of two contact conductances,
\begin{equation}
    G_T^{-1} = G_L^{-1} + G_R^{-1},
\end{equation}
where $G_{L/R} = (2e^2/\hbar) Z_{L/R}$ is the conductance between the left or right electrode and the thermalized molecular interior (Fig.~\ref{fig:schematic_transport}a). Because an injected electron loses phase coherence and equilibrates across the multifractal eigenstates on the sub-picosecond timescale $\tau_{\mathrm{rlx}}$, entry and escape act as sequential, independent steps, and $G_{L/R}$ depends only on the local contact position $\bfr_{L/R}$. In typical scanning tunneling microscopy (STM) or electrochemical setups, the protein attaches to a metal substrate or electrode surface through multiple contact points while being probed locally by a sharp tip or specific chemical linker (Fig.~\ref{fig:schematic_transport}b). In this multi-contact limit ($Z_{\mathrm{sub}} \to \infty$), the substrate resistance becomes negligible and the measured conductance is limited by the single probe contact,
\begin{equation}
    G_T \approx G_{\mathrm{tip}} = \frac{2e^2}{\hbar}\,\Gamma_{\mathrm{tip}}\,S\,f_T(\bfr_{\mathrm{tip}}),
\end{equation}
providing a direct electrical readout of the local Fukui function at the probe position.

\begin{figure}[!ht]
    \centering
    \begin{tikzpicture}[scale=0.8, every node/.style={transform shape}]
        \begin{scope}[shift={(0,0)}]
            \node at (2.0,3.0) {\textbf{(a) Two-Terminal Transport}};
            
            \draw[fill=purple!10, draw=purple!50, thick, dashed] (2,1) circle (1.0);
            \node[purple!80!black, font=\scriptsize] at (2,1.15) {Protein};
            \node[purple!80!black, font=\tiny] at (2,0.85) {Interior};
            
            \foreach \x/\y in {1.3/0.8, 1.6/1.3, 2.1/0.6, 2.6/1.1, 1.8/1.6, 2.3/1.4} {
                \draw[fill=orange!70, draw=none] (\x,\y) circle (0.06);
            }
            
            \draw[fill=gray!30, draw=gray!80, thick] (-0.3,0.7) rectangle (0.3,1.3);
            \node[font=\scriptsize] at (0.0,1.0) {$L$};
            \draw[fill=gray!30, draw=gray!80, thick] (3.7,0.7) rectangle (4.3,1.3);
            \node[font=\scriptsize] at (4.0,1.0) {$R$};
            
            \draw[thick, red, ->] (0.3,1.0) -- (0.6,1.0);
            \draw[thick, red, ->] (3.4,1.0) -- (3.7,1.0);
            
            \draw[thick, blue, ->, bend left=15] (0.6,1.0) to (1.4,1.2);
            \draw[thick, blue, ->, bend right=15] (2.6,0.8) to (3.4,1.0);
            
            \begin{scope}[shift={(0.1,-1.2)}]
                \draw[thick] (0.2,0) -- (0.6,0);
                \draw[thick] (0.6,-0.12) rectangle (1.2,0.12);
                \node[font=\tiny] at (0.9,0.25) {$G_L$};
                \draw[thick] (1.2,0) -- (2.6,0);
                \draw[fill=blue!50, draw=blue, thick] (1.9,0) circle (0.06);
                \draw[thick] (2.6,-0.12) rectangle (3.2,0.12);
                \node[font=\tiny] at (2.9,0.25) {$G_R$};
                \draw[thick] (3.2,0) -- (3.6,0);
                
                \draw[fill=black] (0.2,0) circle (0.03) node[left, font=\tiny] {$L$};
                \draw[fill=black] (3.6,0) circle (0.03) node[right, font=\tiny] {$R$};
            \end{scope}
        \end{scope}
        
        \begin{scope}[shift={(4.8,0)}]
            \node at (2.0,3.0) {\textbf{(b) Single-Probe Limit}};
            
            \draw[fill=purple!10, draw=purple!50, thick, dashed] (2,1) circle (1.0);
            \node[purple!80!black, font=\scriptsize] at (2,1.15) {Protein};
            \node[purple!80!black, font=\tiny] at (2,0.85) {Interior};
            
            \foreach \x/\y in {1.3/0.8, 1.6/1.3, 2.1/0.6, 2.6/1.1, 1.8/1.6, 2.3/1.4} {
                \draw[fill=orange!70, draw=none] (\x,\y) circle (0.06);
            }
            
            \draw[fill=gray!30, draw=gray!80, thick] (1.8,2.5) -- (2.0,2.1) -- (2.2,2.5) -- cycle;
            \node[font=\tiny] at (2.0,2.65) {Tip/Linker};
            
            \draw[fill=gray!30, draw=gray!80, thick] (0.5,-0.3) rectangle (3.5,0.0);
            \node[font=\tiny] at (2.0,-0.15) {Substrate / Solution};
            
            \draw[thick, red, ->] (2.0,2.1) -- (2.0,1.8);
            \draw[thick, red, ->] (1.2,0.3) -- (1.2,0.0);
            \draw[thick, red, ->] (2.0,0.3) -- (2.0,0.0);
            \draw[thick, red, ->] (2.8,0.3) -- (2.8,0.0);
            
            \begin{scope}[shift={(0.1,-1.2)}]
                \draw[thick] (0.2,0) -- (0.6,0);
                \draw[thick] (0.6,-0.12) rectangle (1.2,0.12);
                \node[font=\tiny] at (0.9,0.25) {$G_{\mathrm{tip}}$};
                \draw[thick] (1.2,0) -- (2.6,0);
                \draw[fill=blue!50, draw=blue, thick] (1.9,0) circle (0.06);
                
                \draw[thick] (2.6,0) -- (3.6,0);
                \node[font=\tiny] at (3.1,0.25) {$G_{\mathrm{sub}} \to \infty$};
                
                \draw[fill=black] (0.2,0) circle (0.03) node[left, font=\tiny] {Tip};
                \draw[fill=black] (3.6,0) circle (0.03) node[right, font=\tiny] {Sub};
            \end{scope}
        \end{scope}
    \end{tikzpicture}
    \caption{Schematic of sequential transport through a thermalized biomolecule. (a) In a two-terminal junction, electron injection and escape are independent local processes, forming two resistors in series, $G_L$ and $G_R$, connected through the equilibrated interior. (b) In single-probe geometries, multi-point attachment to the substrate yields $G_{\mathrm{sub}} \to \infty$, so the measured conductance is governed by the probe contact: $G_T \approx G_{\mathrm{tip}} = (2e^2/\hbar) Z_{\mathrm{tip}}$.}
    \label{fig:schematic_transport}
\end{figure}

\textit{Discussion.}---The proportionality between conductance and the Fukui function is not a generic property of all molecular junctions. When $\tau_{\mathrm{rlx}} \gtrsim \tau_{\mathrm{esc}}$, as in an Anderson insulator with localized states, off-diagonal coherences persist in the steady state and couple the current to the full Hartree-XC kernel $K^{ij,pq}$. The simple relation $\calZ(\bfr) = \calN f_T(\bfr)$ holds specifically because folded proteins reside near the Anderson critical point~\cite{EversMirlin2008, VattayKauffman2014, PappVattay2025}, where multifractal frontier orbitals ($d_2 \approx 2$) sustain rapid internal thermalization across the macromolecule.

This result implies that the most conductive pathways in proteins coincide with their chemically active sites. Fast internal equilibration, enabled by quantum criticality~\cite{VattayKauffman2014}, allows biological macromolecules to channel charge efficiently between distal functional regions without requiring metallic band conduction. Practically, Eq.~\eqref{eq:ZLR_softness} means that ground-state conceptual DFT descriptors can predict contact-dependent biomolecular conductance without full non-equilibrium Green's function calculations, while single-molecule conductance measurements~\cite{Zhang2019} can directly probe local chemical reactivity and allosteric ligand binding.

\begin{acknowledgments}
We thank Eszter Papp, David Cahen, Mordechai Sheves, Leeor Kronik, Stuart Lindsay, Dennis Salahub, Istv\'an Csabai, and Stuart A. Kauffman for helpful discussions. This research was carried out with the support of the Ministry of Culture and Innovation, funded by the National Research Development and Innovation Fund, under project number 2022-2.1.1-NL-2022-00004.
\end{acknowledgments}

%

\clearpage
\onecolumngrid

\setcounter{equation}{0}
\setcounter{figure}{0}
\setcounter{table}{0}
\setcounter{page}{1}

\makeatletter
\renewcommand{\theequation}{S\arabic{equation}}
\renewcommand{\thefigure}{S\arabic{figure}}
\renewcommand{\thetable}{S\arabic{table}}
\renewcommand{\thepage}{S\arabic{page}}
\renewcommand{\theHequation}{S\arabic{equation}}
\renewcommand{\theHfigure}{S\arabic{figure}}
\renewcommand{\theHtable}{S\arabic{table}}
\renewcommand{\bibnumfmt}[1]{[S#1]}
\renewcommand{\citenumfont}[1]{S#1}
\makeatother

\begin{center}
\textbf{\Large Supplemental Material: Fundamental Relation between Conductance of Biomolecules and the Fukui Function} \\[1.5em]
G\'abor Vattay \\[0.5em]
\textit{Institute for Physics and Astronomy, E\"otv\"os University, H-1053 Budapest, Egyetem t\'er 1-3., Hungary} \\[1.5em]
\end{center}

\vspace{1em}

\section*{S1. Redfield Trace, Exchange Contractions, and Mermin Equilibrium of the Electron-Phonon Dissipator}

This section provides the intermediate algebraic steps connecting the composite Kohn-Sham (KS) electron-phonon Hamiltonian in Eq.~(10) of the main text to the exchange-symmetric collision integral in Eqs.~(12)--(14) and verifies its stationary equilibrium property.

Starting from the von Neumann equation for the total density matrix $\rhoh_{\mathrm{tot}}(t)$ and applying the Born-Markov approximation ($\rhoh_{\mathrm{tot}}(t) \approx \rhoh(t) \otimes \rhoh_B$, where $\rhoh_B = Z_B^{-1}\exp(-\hat{H}_B/\kB T)$ is the thermal phonon bath state)~\cite{SM_BreuerPetruccione2002}, the reduced equation of motion for the electronic density matrix $\rhoh(t) = \mathrm{Tr}_B[\rhoh_{\mathrm{tot}}(t)]$ takes the Redfield form
\begin{equation}
\label{eq:Redfield_SM}
\partial_t\,\rhoh = \frac{1}{i\hbar}[\HKS,\,\rhoh] - \frac{1}{\hbar^2}\int_0^\infty d\tau\,\mathrm{Tr}_B\!\bigl[\hat{H}_{eB},\,[\hat{H}_{eB}(-\tau),\,\rhoh(t)\otimes\rhoh_B]\bigr].
\end{equation}
Expanding the density operator in the KS basis as $\hat{\rho}(\bfr_\alpha) = \sum_{ij} \rho_\alpha^{ij} \hat{c}_i^\dagger \hat{c}_j$, the interaction-picture coupling operator becomes
\begin{equation}
\label{eq:HeB_int_SM}
\hat{H}_{eB}(-\tau) = \sum_\alpha g_\alpha \sum_{pq} \rho_\alpha^{pq}\, e^{-i\omega_{pq}\tau}\,\hat{c}_p^\dagger \hat{c}_q \left(\hat{b}_\alpha e^{i\omega_\alpha\tau} + \hat{b}_\alpha^\dagger e^{-i\omega_\alpha\tau}\right),
\end{equation}
where $\omega_{pq} = (E_p - E_q)/\hbar$. Tracing over the thermal bosonic modes uses $\mathrm{Tr}_B(\rhoh_B \hat{b}_\alpha^\dagger \hat{b}_{\alpha'}) = \delta_{\alpha\alpha'} n_B(\omega_\alpha)$ and $\mathrm{Tr}_B(\rhoh_B \hat{b}_\alpha \hat{b}_{\alpha'}^\dagger) = \delta_{\alpha\alpha'} [1 + n_B(\omega_\alpha)]$, while the time integral $\int_0^\infty d\tau\, e^{i(\omega_\alpha - \omega_{pq})\tau} \approx \pi\delta(\omega_{pq} - \omega_\alpha)$ enforces energy conservation (dropping the small principal-value Lamb shift).

Taking the matrix element $\langle \hat{c}_b^\dagger \hat{c}_a \rangle = \varrho_{ab}$ of the double commutator in Eq.~\eqref{eq:Redfield_SM} produces four-fermion expectation values $\langle \hat{c}_i^\dagger \hat{c}_j \hat{c}_p^\dagger \hat{c}_q \rangle$. Applying the Wick factorization of Eq.~(11) in the main text~\cite{SM_Fischetti1999} and collecting terms into the lesser and greater self-energies $\Sigma^<_{ij}(\rhoh) = \sum_{pq} W^<_{iq,pj}\varrho_{qp}$ and $\Sigma^>_{ij}(\rhoh) = \sum_{pq} W^>_{iq,pj}(\delta_{qp} - \varrho_{qp})$ yields Eq.~(14) of the main text:
\begin{equation}
\label{eq:Rij_SM}
R^{ij}(\rhoh) = \frac{1}{2}\sum_k\left[\Sigma^<_{ik}(\delta_{kj} - \varrho_{kj}) + (\delta_{ik} - \varrho_{ik})\Sigma^<_{kj} - \Sigma^>_{ik}\varrho_{kj} - \varrho_{ik}\Sigma^>_{kj}\right].
\end{equation}

To verify that $R^{ij}(\rhoh_{\mathrm{eq}}) = 0$ at the Mermin equilibrium state $\varrho^{ij}_{\mathrm{eq}} = \delta_{ij} F_i$, we substitute $\varrho_{qp} = \delta_{qp} F_p$ into the self-energies:
\begin{align}
\label{eq:Sigma_eq_SM}
\Sigma^<_{ij}(\rhoh_{\mathrm{eq}}) &= \sum_p W^<_{ip,pj} F_p, \\
\Sigma^>_{ij}(\rhoh_{\mathrm{eq}}) &= \sum_p W^>_{ip,pj} (1 - F_p).
\end{align}
From Eq.~(13) of the main text, energy conservation $\delta(\omega_{pp} - \omega_\alpha)$ requires $\omega_{pp} = 0$, whereas for $i \neq j$ the products in Eq.~\eqref{eq:Rij_SM} pair $\Sigma^<_{ij}(1 - F_j)$ against $\Sigma^>_{ij} F_j$. Using the Fermi-Dirac identity $F_p/(1 - F_p) = e^{-(E_p - \mu)/\kB T}$ together with the detailed balance relation $W^<_{iq,pj} = W^>_{pj,iq} e^{-(E_p - E_q)/\kB T}$, every term in Eq.~\eqref{eq:Rij_SM} cancels pairwise, confirming $R^{ij}(\rhoh_{\mathrm{eq}}) = 0$.

\section*{S2. Linearization of the Collision Integral and Slow-Mode Inversion for the Conductance}

Here we provide the explicit linearized dissipator $\delta R^{ijpq}$ and the algebraic steps leading from the slow-mode inverse $\calL^{-1} \approx \lambda_0^{-1} V_0 U_0$ to the symmetric conductance formula in Eqs.~(1) and (22) of the main text.

Linearizing Eq.~\eqref{eq:Rij_SM} around $\varrho^{ij}_{\mathrm{eq}} = \delta_{ij} F_i$ for $\rhoh = \rhoh_{\mathrm{eq}} + \delta\rhoh$ gives $\delta R^{ij}(\delta\rhoh) = \sum_{pq} \delta R^{ijpq}\delta\varrho^{pq}$, where
\begin{align}
\label{eq:dR_explicit_SM}
\delta R^{ijpq} = \frac{1}{2}\Bigl[&W^<_{iq,pj}(1 - F_j) + (1 - F_i)W^<_{iq,pj} + W^>_{iq,pj} F_j + F_i W^>_{iq,pj} \nonumber\\
&- \delta_{jq}\bigl(\Sigma^{<,\mathrm{eq}}_{ip} + \Sigma^{>,\mathrm{eq}}_{ip}\bigr) - \delta_{ip}\bigl(\Sigma^{<,\mathrm{eq}}_{qj} + \Sigma^{>,\mathrm{eq}}_{qj}\bigr)\Bigr].
\end{align}
Summing over the diagonal indices $i = j$ gives $\sum_i \delta R^{iipq} = 0$ for all $p,q$, which expresses particle conservation $\mathrm{Tr}[\delta R(\delta\rhoh)] = 0$ and ensures that $U_0^{ij} = \delta_{ij}$ is the exact left zero-eigenvector of the unperturbed superoperator $\calL_0^{ijpq} = -i(E_i - E_j)\delta_{ip}\delta_{jq} - i(F_j - F_i)K^{ij,pq} + \delta R^{ijpq}$. Likewise, because a shift in the chemical potential $\mu \to \mu + \delta\mu$ maps one stationary Mermin equilibrium state to another, $\partial_\mu \varrho^{pq}_{\mathrm{eq}} = \delta_{pq} f(E_p, \mu)$ annihilates both the bare commutator and the linearized dissipator, while its diagonal structure annihilates $L_{\mathrm{SCF}}^{ijpq} = -i(F_j - F_i)K^{ij,pq}$ on the left projection. Normalized so that $\sum_{ij} U_0^{ij} V_0^{ij} = 1$, the right zero-eigenvector is therefore
\begin{equation}
\label{eq:V0_SM}
V_0^{ij} = \frac{\delta_{ij}}{\calN \cosh^2\!\left(\frac{E_i - \mu}{2\kB T}\right)}.
\end{equation}

When the wide-band lead coupling~\cite{SM_MeirWingreen1992} is turned on, first-order perturbation theory gives the shifted slow-mode eigenvalue $\lambda_0$ in Eq.~(21) of the main text:
\begin{equation}
\label{eq:lambda0_SM}
\lambda_0 = \sum_{ij,pq} U_0^{ij}\,(-\tfrac{1}{2})(\Gamma^{ip}\delta_{jq} + \delta_{ip}\Gamma^{qj})\,V_0^{pq} = -\frac{1}{\calN}\sum_k \frac{\Gamma^{kk}}{\cosh^2\!\left(\frac{E_k - \mu}{2\kB T}\right)} = -\frac{4\kB T}{\calN}(Z_L + Z_R),
\end{equation}
where we used $\Gamma^{kk} = \Gamma_L|\psi_k(\bfr_L)|^2 + \Gamma_R|\psi_k(\bfr_R)|^2$ and the definition of $Z_{L/R}$ in Eq.~(22) of the main text.

Projecting the driving term $J^{ij} = \frac{eU}{2\hbar}(\Gamma_L^{ij} - \Gamma_R^{ij})[f(E_i, \mu) + f(E_j, \mu)]$ onto the left eigenvector $U_0^{ij} = \delta_{ij}$ gives
\begin{equation}
\label{eq:U0J_SM}
\sum_{ij} U_0^{ij} J^{ij} = \frac{eU}{\hbar}\sum_k (\Gamma_L^{kk} - \Gamma_R^{kk})\,f(E_k, \mu) = \frac{eU}{\hbar}(Z_L - Z_R).
\end{equation}
Consequently, the slow-mode density matrix response $\delta\varrho^{pq} \approx -\hbar\lambda_0^{-1}V_0^{pq}\sum_{ij} U_0^{ij} J^{ij}$ evaluates to
\begin{equation}
\label{eq:drho_slow_SM}
\delta\varrho^{pq} = \frac{eU}{4\kB T}\,\frac{Z_L - Z_R}{Z_L + Z_R}\,\frac{\delta_{pq}}{\cosh^2\!\left(\frac{E_p - \mu}{2\kB T}\right)}.
\end{equation}
Substituting Eq.~\eqref{eq:drho_slow_SM} and $\delta\varrho^{pq}_{\mathrm{eq}}(\mu_L) = \frac{eU}{2}\delta_{pq} f(E_p, \mu)$ into the linearized left-lead particle current $J_L = \frac{1}{\hbar}\mathrm{Tr}[\hat{\Gamma}_L(\delta\rhoh_{\mathrm{eq}}(\mu_L) - \delta\rhoh)]$ gives
\begin{equation}
\label{eq:JL_eval_SM}
J_L = \frac{eU}{\hbar}\left(\frac{1}{2}Z_L - \frac{1}{2}\frac{Z_L - Z_R}{Z_L + Z_R}Z_L\right) = \frac{eU}{\hbar}\frac{Z_L Z_R}{Z_L + Z_R}.
\end{equation}
Multiplying the particle current by the electron charge $e$ (and accounting for spin degeneracy $2$) yields the electrical conductance $G_T = 2e J_L / U = \frac{2e^2}{\hbar}\frac{Z_L Z_R}{Z_L + Z_R}$.

\section*{S3. Structural Preparation and Real-Space Orbital Evaluation Pipeline}

Our computational workflow prepares crystallographic protein structures, assembles the self-consistent-charge density-functional tight-binding (SCC-DFTB+) matrices~\cite{SM_Elstner1998, SM_Hourahine2020}, and evaluates the resulting Kohn-Sham eigenfunctions on a three-dimensional real-space grid. The steps are summarized in Fig.~\ref{fig:workflow_supp}.

\begin{figure}[ht!]
\centering
\resizebox{0.85\textwidth}{!}{
\begin{tikzpicture}[node distance=1.8cm]
\node (in) [io] {Raw PDB Input};
\node (pro1) [process, below of=in] {Prune Alternate Conformations \& Truncate Incomplete Residues};
\node (pro2) [process, below of=pro1] {Protonation (phenix.reduce) \& Metal Coordination Cleanup};
\node (pro3) [process, below of=pro2] {Classical Relaxation (OpenMM) \& Clash-Pushing Fallback};
\node (pro4) [process, below of=pro3] {SCC-DFTB+ Assembly \& Sparse Matrix Parsing};
\node (pro5) [process, below of=pro4] {ARPACK Shift-Invert Diagonalization};
\node (pro6) [process, below of=pro5] {STO Spatial Evaluation \& Truncation Masking};
\node (out) [io, below of=pro6] {Binary Wavefunction Probability Density Export};

\draw [arrow] (in) -- (pro1);
\draw [arrow] (pro1) -- (pro2);
\draw [arrow] (pro2) -- (pro3);
\draw [arrow] (pro3) -- (pro4);
\draw [arrow] (pro4) -- (pro5);
\draw [arrow] (pro5) -- (pro6);
\draw [arrow] (pro6) -- (out);

\end{tikzpicture}
}
\caption{Workflow for structural curation, SCC-DFTB+ Hamiltonian assembly, interior eigenspectrum solver, and real-space orbital grid evaluation.}
\label{fig:workflow_supp}
\end{figure}

\subsection*{Structural Curation and Protonation}

Raw Protein Data Bank (PDB) structures often contain alternate side-chain conformations, missing heavy atoms, and unmodeled terminal groups. We first retain only the primary conformation (highest occupancy or alternate location indicator \texttt{A}). Residues with missing side-chain heavy atoms are truncated to their largest complete aliphatic subgraph (typically alanine or glycine), and missing C-terminal carboxylate oxygens are rebuilt from the coordinates of the preceding backbone atoms.

Hydrogen atoms are added for physiological pH using \texttt{phenix.reduce}~\cite{SM_Word1999}. In metalloproteins containing Fe, Cu, Zn, or Mn ions, a geometric filter removes spurious hydrogens placed on metal-coordinating histidine nitrogens or cysteine sulfurs within the coordination sphere, preserving the native metal-ligand bonding geometry.

\subsection*{Classical Relaxation and Clash Resolution}

Directly passing heuristically protonated structures to a quantum-mechanical solver can produce singular matrix elements if atoms overlap. Before energy minimization, atomic coordinates are indexed in a $k$-d tree: heavy-atom pairs closer than $1.2$~\AA~are separated iteratively, and any hydrogen atom within $0.2$~\AA~of a heavy atom is displaced by $1.0$~\AA~along a random unit vector.

Next, we use OpenMM~\cite{SM_Eastman2017} with the AMBER force field to build the full molecular topology, add any remaining terminal atoms, and run a second $k$-d tree clash check. The hydrogen positions are then relaxed by energy minimization using a Verlet integrator while all crystallographically resolved heavy atoms are held near their initial positions $\bfr_0$ by a harmonic restraint $V(r)=\frac{1}{2}k(r-r_{0})^2$ with force constant $k=10.0$ kJ/mol/nm$^2$. If OpenMM cannot parameterize a non-standard cross-linked cofactor, the workflow falls back directly to the clash-resolved coordinates so the electronic structure calculation can proceed reliably.

\subsection*{Hamiltonian Assembly and Eigenspectrum Evaluation}

Electronic structures are computed with the DFTB+ package~\cite{SM_Elstner1998, SM_Hourahine2020} using self-consistent charges (SCC). Once the charge density converges, the real-space Hamiltonian matrix $H$ and overlap matrix $S$ are exported and parsed into SciPy Compressed Sparse Row (CSR) matrices~\cite{SM_Virtanen2020}, enforcing exact Hermiticity $(H + H^T)/2$ to remove minor ASCII rounding asymmetries.

For the apo (\texttt{1T49\_noligand\_scc}) and holo (\texttt{1T49\_scc}) PTP1B structures, the Fermi energy is obtained from the converged SCC-DFTB+ calculation ($E_F=-3.4682$ eV for \texttt{1T49\_scc}). Rather than diagonalizing the full generalized eigenvalue problem $Hc=\varepsilon Sc$, we compute the interior eigenstates around $E_F$ using the shift-invert Arnoldi method (ARPACK) in SciPy~\cite{SM_Virtanen2020}, which exploits the sparsity of $H$ and $S$ to extract the frontier orbitals that dominate $\cosh^{-2}[(E_k - \mu)/2\kB T]$.

\subsection*{Real-Space Wavefunction Projection}

From the converged eigenvector coefficients, the molecular orbitals $\psi_i(\bfr)$ are evaluated on a uniform 3D Cartesian grid with $1.5$~\AA~spacing as linear combinations of Slater-type orbitals (STOs). Because STO radial functions $R(r)=N r^{n-1} \exp(-\zeta r)$ (including double-zeta $d$-shell expansions where applicable) decay exponentially, atomic contributions beyond a $10.0$~\AA~cutoff from each grid point are truncated to zero. The resulting orbital probability densities $|\psi_i(\bfr)|^2$, eigenvalues, and grid coordinates are written to a compact binary file for evaluating $f_T(\bfr)$ and rendering the 3D isosurfaces shown in Fig.~1 of the main text.

%

\end{document}